\documentclass[twocolumn,showpacs,preprintnumbers,amsmath,amssymb]{revtex4}
\usepackage{graphicx}
\usepackage{dcolumn}
\usepackage{bm}
\def\bea {\begin{eqnarray}}
\def\eea {\end{eqnarray}}

\def\be {\begin{equation}}
\def\ee {\end{equation}}
\def\nn {\nonumber}
\begin{document}

\title{Probing quark gluon plasma properties by heavy flavours}

\author{Santosh K Das, Jan-e Alam and Payal Mohanty}

\medskip

\affiliation{Variable Energy Cyclotron Centre, 1/AF, Bidhan Nagar , 
Kolkata - 700064}

\date{\today}
\begin{abstract}
The Fokker Planck (FP) equation has been solved to study the interaction
of non-equilibrated heavy quarks  with the Quark Gluon Plasma (QGP) expected 
to be formed in heavy ion collisions at RHIC energies.
The solutions of the FP equation have been convoluted with the relevant 
fragmentation functions to obtain the $D$ and $B$ meson spectra. The results are compared 
with experimental data measured by STAR collaboration. It is found that the present 
experimental data can not distinguish between the $p_T$ spectra obtained from the 
equilibrium and non-equilibrium charm distributions. Data at lower $p_T$ may play a 
crucial role in making the distinction between the two.  The nuclear suppression 
factor, $R_{\mathrm AA}$ for non-photonic single electron spectra resulting from 
the semileptonic decays of hadrons containing heavy flavours have been evaluated 
using the present formalism.  It is observed that the experimental data on nuclear 
suppression factor of the non-photonic electrons can be reproduced within this 
formalism by enhancing the pQCD cross sections by a factor of 2 
provided the expansion of the bulk matter is governed by the velocity of 
sound, $c_s\sim 1/\sqrt{4}$.  Ideal gas equation of state fails to reproduce the 
data even with the enhancement of the pQCD cross sections by a factor of 2.

\end{abstract}

\pacs{12.38.Mh,25.75.-q,24.85.+p,25.75.Nq}
\maketitle

\section{Introduction}

The nuclear collisions at Relativistic Heavy Ion Collider (RHIC) 
and the Large Hadron Collider (LHC) energies are aimed at creating 
a phase of matter where the properties of the matter is governed 
by quarks and gluons~\cite{qm08}. Such a phase of matter, composed of mainly
light quarks and gluons - is called quark gluon
plasma (QGP). The study of the bulk properties of QGP is a 
field of high contemporary interest and the heavy flavours namely, charm
and bottom quarks play a crucial role in such studies, because they are
produced in the early stage of the collisions, they are not part of 
the bulk properties of the system and their thermalization time scale
is larger than the light quarks and gluons and hence can retain the
interaction history more effectively.   

The successes of the relativistic hydrodynamical 
model~\cite{pasi,teaney} 
in describing the host of experimental results from RHIC~\cite{npa2005}
indicate that the thermalization might have taken place in the system
of quarks and gluons formed after the nuclear collisions. 
The strong final state interaction of high energy partons  with
the QGP {\it i.e.} the observed jet quenching~\cite{phenix1,star1} and
the large elliptic flow ($v_2$)~\cite{phenix2,star2} in Au+Au collisions at RHIC
indicate the possibility of fast equilibration.
On the one hand 
the experimental data indicate early thermalization time $\sim 0.6$ fm/c
~\cite{arnold1} on the other hand the pQCD based calculations give a 
thermalization time $\sim 2.5$ fm/c~\cite{baier}(see also~{\cite{raju}). 
The gap between
these two time scales suggests that  the non-perturbative effects 
play a crucial role in achieving thermalization. It has also been
pointed out that the instabilities~\cite{mrowczynski,romatschke,arnold2,
arnold3} may derive the system towards faster equilibrium.
The two pertinent issues regarding the equilibration 
which will be addresses here
are (i) do the the heavy quarks 
achieve equilibrium and (ii) in case they achieve equilibration,
can the equilibrium be maintained during expansion of the system. 
The second issue will be addressed first.

We make a rather strong assumptions that the heavy quarks 
produced initially 
is in thermal equilibrium and check whether it can 
maintain the equilibrium during the entire evolution processes by
comparing their scattering rates with the expansion rate of the matter.
This issue will be addressed with different
equation of states (EoS) which affects the expansion rate.
In case the heavy quarks are unable to maintain the equilibrium then
the analysis of the transverse momentum of the mesons carrying heavy
flavours can not be done by using thermal phase space distribution.
The analysis will require non-equilibrium statistical mechanical treatment.
We solve Fokker-Planck (FP) 
equation~\cite{sc,svetitsky,japrl,moore,rapp,turbide,bjoraker,munshi}
to address this issue as discussed below.

The perturbative QCD (pQCD) calculations indicate that the heavy quark
thermalization time, $\tau_i^Q$ is larger~\cite{moore} than the light quarks 
and gluons thermalization scale $\tau_i$.
Gluons may thermalize before 
up and down quarks~\cite{japrl,shuryak}, in the present work we 
assume that the QGP is formed at time $\tau_i$. Therefore, the interaction
of the non-equilibrated heavy quarks ($Q$) with the equilibrated QGP for
the time interval $\tau_i<\tau<\tau_i^Q$ can be treated within the ambit 
of the FP equation {\it i.e.} the heavy quark can be thought of
executing Brownian motion in the heat bath of QGP during the said interval
of time. The solution of the FP equation can be used
to study $p_T$ spectra of heavy mesons in the spirit of blast wave 
method.

In the next section we address the issues of thermalization in a
rapidly expanding system. The results indicate that the heavy quark
can not maintain the equilibrium at RHIC and LHC energies during the
entire evolution history of the QGP. This demands the treatment of 
the problem within the framework of non-equilibrium statistical 
mechanics, which is discussed in section III. Section IV 
is devoted  to summary and conclusions.

\section{Thermalization in an expanding system}
We consider a thermally equilibrated partonic system of quarks, anti-quarks and gluons 
produced in relativistic heavy ion collisions. We would like to study whether the
system can maintain thermal equilibrium when it evolves in space and time. 
Relativistic hydrodynamics (with boost invariance along longitudinal direction
and cylindrical symmetry) have been used to describe the space-time evolution.
For this purpose, the scattering time scale ($\tau_{\mathrm scatt}$) of the partons are compared
with the expansion time scale ($\tau_{\mathrm exp}$). For maintenance of thermal equilibrium
the following criteria should be satisfied:  
\be
\tau_{\mathrm exp}\geq\alpha\tau_{\mathrm scatt}
\label{Eq1}
\ee
where $\alpha\sim O(1)$ is a constant. 
The criteria given in Eq.~\ref{scattvsexp} is reverse to the
one used to study the freeze-out of various species 
of particles during the evolution of the early 
universe~\cite{kolbandturner} (similar condition is used in ~\cite{navarra}
in heavy ion collisions also). 

The $\tau_{\mathrm scatt}$ is determined for each partons by the expression
\be
\tau_{\mathrm scatt}^i=\frac{1}{\sum\sigma_{ij}v_{ij}n_j}
\label{scattvsexp}
\ee
where $\sigma_{\mathrm ij}$ is the total cross section for  
particles $i$ and $j$, $v_{ij}$ is the relative velocity between
the particles $i$ and $j$ and $n_j$ is the density of the particle type $j$.

\par
To calculate the scattering time we use the following 
 processes $gg \rightarrow gg$,
$gg \rightarrow q\overline{q}$, 
$q(\overline{q})g \rightarrow q(\overline{q})g$, 
$qq \rightarrow qq$, 
$q\overline{q} \rightarrow q\overline{q}$ for 
light flavours and gluons~\cite{pqcd}.
Here $q$ stands for light quarks and $g$ denotes
gluons. For evaluating $\tau_{\mathrm scatt}$ for 
heavy quarks ($Q$) the pQCD processes
are taken from ~\cite{combridge}.
The infrared divergence appearing in case of massless particle
exchange in the $t$-channel has been shielded by Debye mass.

The expansion time scale can be defined as:
\be
\tau_{\mathrm exp}^{-1}=\frac{1}{\epsilon(\tau,r)}
\frac{d\epsilon(\tau,r)}{d\tau}
\ee
where $\epsilon(\tau,r)$ is the energy density, $\tau$ and $r$ are
proper time and the radial co-ordinate respectively. $\epsilon(\tau,r)$
is calculated by solving the hydrodynamical equation:
\be
\partial_\mu T^{\mu\nu}=0
\label{hydro}
\ee
with the assumption of boost invariance along longitudinal
direction~\cite{bjorken} and cylindrical symmetry of the system~\cite{hvg}. 
In Eq.~(\ref{hydro}), $T^{\mu\nu}=(\epsilon+P)u^\mu u^\nu-g^{\mu\nu}P$ is
the energy momentum tensor, $P$ is the pressure, $u^\mu$ denotes
four velocity and $g^{\mu\nu}$ stands for metric tensor. We consider
a net baryon free QGP here, therefore the baryonic chemical potential ($\mu_B$)
is zero.
 
The expansion rates for RHIC and LHC energies have been calculated using  the 
initial conditions, $T_i=400$ MeV, $\tau_i=0.2$ fm  for RHIC which gives
$dN/dy\sim 1100$~\cite{npa2005} and
$T_i=700$ MeV, $\tau_i=0.08$ fm for LHC giving $dN/dy=2100$~\cite{lhc}. 
The initial radial velocity has been
taken as zero for both the cases. Two sets (SET-I and SET-II) 
of equation of state (EoS) have been
used to study the sensitivity of the results on EoS. 

SET-I: In a first order phase transition scenario -
we use the bag model EOS for the QGP phase and for the hadronic
phase all the resonances with mass $\leq 2.5$ GeV have been
considered~\cite{bm}.
and SET-II: The EOS is taken from lattice QCD calculations performed by 
the MILC collaboration~\cite{MILC}.

\begin{figure}[h]
\begin{center}
\includegraphics[scale=0.43]{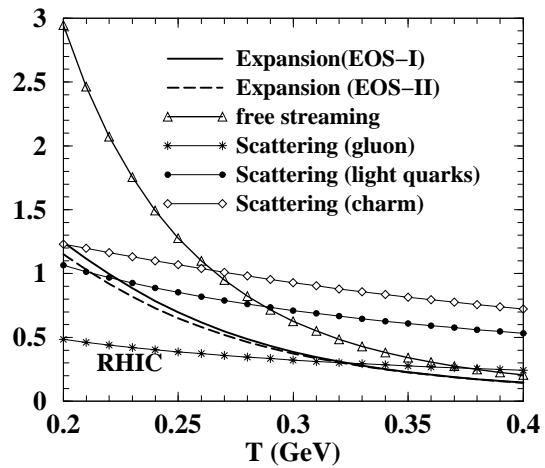}
\caption{Expansion time vs scattering time (calculated with pQCD process) 
in RHIC energy.  Here the expansion time scale has been calculated at $r=1$ fm. 
}
\label{fig1}
\end{center}
\end{figure}

\begin{figure}[h]
\begin{center}
\includegraphics[scale=0.43]{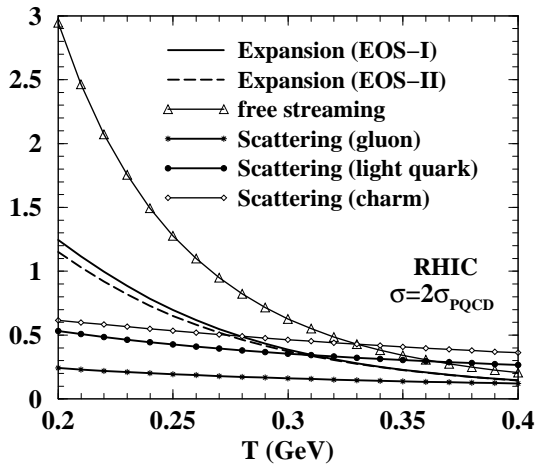}
\caption{Same as Fig.~\protect{\ref{fig1}} with the pQCD cross section 
enhanced by a factor 2}
\label{fig2}
\end{center}
\end{figure}

\par
In Fig~\ref{fig1} the scattering time scale is contrasted with
the expansion time scale for two types of EOS mentioned above. 
For the sake of comparison the expansion rate for the 
extreme case of free streaming
is also displayed. The scattering rates are evaluated with pQCD
cross sections. The condition for equilibration in Eq.~\ref{Eq1}
indicates that the gluons remains close to equilibrium, however the  
charm and bottom (not shown in the figure) quarks remain out of 
equilibrium during the entire evolution history. 

However,  as mentioned in the introduction the analysis of
the experimental data within the ambit of relativistic hydrodynamics
suggest that the matter formed in Au+Au collisions at RHIC achieve
thermalization. One possible reasons for the thermalization to occur
is that the partons interact strongly after their  formation in 
the heavy ion collisions.  
It is argued in~\cite{kovchegov} that the onset of thermalization
in the system formed in heavy ion collisions at relativistic
energies can not be achieved without non-perturbative effects.
It has also been shown in~\cite{molnar} that 
a large enhancement of the pQCD cross section is required for 
the reproduction of experimental data on elliptic flow at RHIC
energies. 
Therefore, the pQCD cross sections used to derive
the results shown in Fig.~\ref{fig1} should include non-perturbative
effects. To implement this we  enhance the pQCD cross sections
by a factor of 2. The resulting scattering time is compared with the 
expansion time in Fig.~\ref{fig2}. It is observed that the gluons
are kept in equilibrium throughout the evolution, light quarks are
closer to the equilibrium as compared to the heavy flavours. 

In Figs.~\ref{fig3}-\ref{fig4} the results for LHC are displayed
for the two time scales mentioned above for pQCD and enhanced
cross sections. The expansion becomes faster at LHC than RHIC 
because of the higher internal pressure. As a consequence,
it is interesting to note that the thermalization scenario at LHC does 
not differ drastically from  RHIC. 


\begin{figure}[h]
\begin{center}
\includegraphics[scale=0.43]{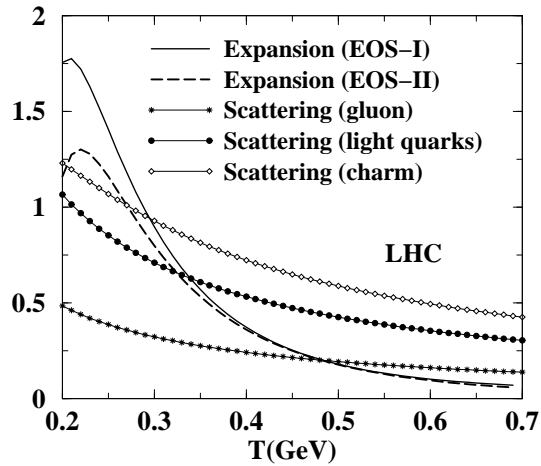}
\caption{Same as Fig.~\protect{\ref{fig1}} for LHC energy. 
}
\label{fig3}
\end{center}
\end{figure}

\begin{figure}[h]
\begin{center}
\includegraphics[scale=0.43]{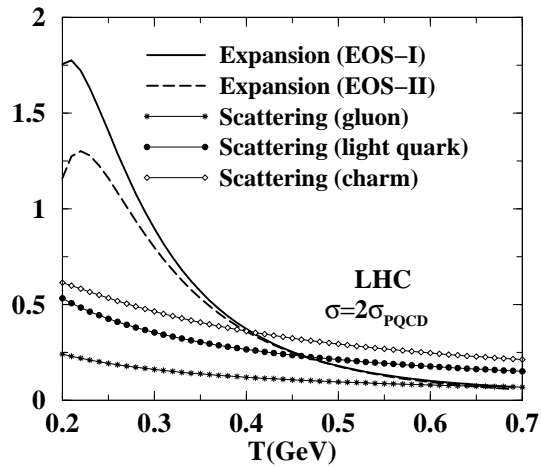}
\caption{Same as Fig.~\protect{\ref{fig2}} for LHC energy. 
}
\label{fig4}
\end{center}
\end{figure}

\section{Non-Equilibrium process}
It is argued in~\cite{moore} that the relaxation time
for heavy quarks is larger than the corresponding quantities
to light partons by a factor of $M/T$, where $M$ is the mass
of the heavy flavour and $T$ is the temperature.
In the present work we have also seen that 
the heavy flavours do not maintain equilibration
throughout the evolution scenario, but
the gluons are close to equilibrium.  Therefore, we treat this
problem as an interaction between equilibrium and non-equilibrium degrees
of freedom and FP equation provides an appropriate framework for such studies.

The Boltzmann transport equation describing a non-equilibrium 
statistical system reads:
\be
\left[\frac{\partial}{\partial t} 
+ \frac{p}{E} \frac{\partial}{\partial x} 
+ F.\frac{\partial}{\partial p}\right]f(x,p,t)=
\left[\frac{\partial f}{\partial t}\right]_{col}
\ee
The assumption of uniformity in the plasma and absence of any external force
leads to
\be 
\frac{\partial f}{\partial t}=
\left[\frac{\partial f}{\partial t}\right]_{col}
\ee
The collision term on the right hand side of the above equation can be 
approximated as (see ~\cite{svetitsky,npa1997} for details): 
\be
\left[\frac{\partial f}{\partial t}\right]_{col} = 
\frac{\partial}{\partial p_i} \left[ A_i(p)f + 
\frac{\partial}{\partial p_i} \lbrack B_{ij}(p) f \rbrack\right] 
\label{expeq}
\ee
where we have defined the kernels 
\begin{eqnarray}
&& A_i = \int d^3 p \omega (p,k) k_i \nonumber\\
&&B_{ij} = \int d^3 p \omega (p,k) k_ik_j.
\end{eqnarray}
and the function $\omega(p,k)$ is given by
\be
\omega(p,k)=g_j\int\frac{d^3q}{(2\pi)^3}f_j(q)v_{ij}\sigma^j_{p,q\rightarrow p-k,q+k}
\ee
where $f_j$ is the phase space distribution for the particle $j$,
$v_{ij}$ is the relative velocity between the two collision partners,
$\sigma$ denotes the cross section and $g_j$ is the statistical
degeneracy. The co-efficients in the first two terms of the expansion
in Eq.~\ref{expeq} are comparable in magnitude because the averaging
of $k_i$ involves greater cancellation than the averaging of the
quadratic term $k_ik_j$. The higher power of $k_i$'s are smaller~\cite{landau}.

With these approximations the Boltzmann equation reduces to a non-linear
integro-differential equation known as Landau 
kinetic equation:
\be
\frac{\partial f}{\partial t} = 
\frac{\partial}{\partial p_i} \left[ A_i(p)f + 
\frac{\partial}{\partial p_i} \lbrack B_{ij}(p) f\rbrack \right] 
\label{landaueq}
\ee
The nonlinearity is caused due to the
appearance of $f$ in $A_i$ and $B_{ij}$ through $w(p,k)$.
It arises from the simple fact that we are studying
a collision process which involves two particles - it should,
therefore, depend on the states of the two participating particles in
the collision process and hence on the product of the two.
Considerable simplicity may be achieved by replacing the distribution
functions of the collision partners of the test particle by their 
equilibrium Fermi-Dirac or Bose-Einstein distributions
(depending on the statistical nature)
in the expressions of $A_i$ and $B_{ij}$. Then Eq.~\ref{landaueq} 
reduces to a linear
partial differential equation - usually referred to as the Fokker-Planck
equation\cite{balescu} describing the motion of a particle which
is out of thermal equilibrium with the particles in a thermal bath.
The quantities $A_i$ and $B_{ij}$ are related to the usual 
drag and diffusion coefficients and we denote them by $\gamma_i$ and
$D_{ij}$ respectively ({\it i.e.} these quantities can be obtained
from the expressions for $A_i$ and $B_{ij}$ by replacing the distribution
functions by their thermal counterparts):
\be
\frac{\partial f}{\partial t} = 
\frac{\partial}{\partial p_i} \left[ \gamma_i(p)f + 
\frac{\partial}{\partial p_i} \lbrack D_{ij}(p) f \rbrack \right] 
\label{FPeq}
\ee
We evaluate the 
value of the $\gamma_i$ and $D_{ij}$ for the reaction: 
gQ $\rightarrow$ gQ and qQ $\rightarrow$ qQ
for both zero and non-zero quark 
chemical potential ($\mu=\mu_B/3$).  
In Fig.~\ref{fig5} we depict the variation of the drag coefficients
as a function of the transverse momentum of the charm and bottom quarks
at a temperature, $T= 200$ MeV.
The momentum dependence is weak. For non-zero quark chemical
potential the  value of the drag increases, however,
the nature of the variations remain same. 
In Fig.~\ref{fig6} the temperature variation of the 
drag co-efficient is plotted for both zero and non-zero quark chemical 
potential. Qualitatively, the inverse of the drag co-efficient gives
the magnitude of the relaxation time. Therefore, the present results
indicate that a system with fixed temperature achieves equilibrium faster for 
non-zero $\mu$. In Fig.~\ref{fig7}  the diffusion coefficients are 
plotted as a function of $p_T$ for $T=200$ MeV. The diffusion
co-efficient for non-zero $\mu$ is larger
as compared to the case of vanishing $\mu$.
The same quantity is displayed in Fig.~\ref{fig8} as a function
of temperature. In the present work we confine to $\mu=0$.
Recently the heavy quark momentum diffusion co-efficient has been
computed~\cite{caron-huot} at next to leading order within the
the ambit of hard thermal loop approximations. For $T\sim 400$ MeV
our momentum averaged
pQCD  value of the diffusion co-efficient is comparable to the value 
obtained in~\cite{caron-huot} in the leading order approximation
for the same set of inputs ({\it e.g.} strong coupling constant, 
number of flavours etc).

The inverse of the drag co-efficients gives an estimate of the thermalization
time scale. Results obtained in the present work indicate that the 
heavy quarks are unlikely to attain thermalization at RHIC and LHC energies
~\cite{mps}.

The total amount of energy dissipated by a parton 
depends on the path length it traverses through the plasma.
Each parton traverse different path length
which depends on the  geometry of the system and on the point 
where its is produced.
The probability that a parton is created at a point $(r,\phi)$
in the plasma depends on the number of binary collisions 
at that point which can be taken as~\cite{turbide}:
\be
P(r,\phi)=\frac{2}{\pi R^2}(1-\frac{r^2}{R^2})\theta(R-r)
\ee
where $R$ is the nuclear radius. 
A parton created at $(r,\phi)$ in the transverse plane
propagate a distance $L=\sqrt{R^2-r^2sin^2\phi}-rcos\phi$
in the medium. In the present work we adopt the following
averaging procedure for the transport coefficients. 
For the drag coefficient ($\gamma$):
\be
\Gamma=\int rdr d\phi P(r,\phi) \int^{L/v}d\tau\gamma(\tau)
\label{cgama}
\ee
where $v$ is the velocity of the propagating partons. The
quantity $\Gamma$ appears in the solution of the FP
equation(see~\cite{rapp} for details). Similar averaging has
been done for the expression involving diffusion co-efficients
to take into account the geometry of the system.

Using the drag and diffusion co-efficients as inputs 
we solve the to FP equation with the following parametrization of the 
initial momentum distribution
of the charm quarks generated in p-p collisions at $\sqrt{s}=200$ GeV:
 \be
\frac{d^2N_c}{dp_T^2} = C \frac{(p_T + A)^2}{(1+\frac{p_T}{B})^\alpha}
\ee
where A=0.5 GeV, B=6.6 GeV, $\alpha=21$ and C=0.845 GeV$^{-4}$. 
We do not elaborate here on the procedure of solving the FP equation
as this has been discussed in detail in ref.~\cite{rapp}.
The corresponding initial distributions for bottom quarks can be obtained
from the results obtained in Ref.~\cite{matteo} 
for pp collisions at $\sqrt{s}=200$ GeV.
Obtaining the solution of the FP equation for the heavy (charm 
and bottom) quarks
we convolute it with the fragmentation functions of the 
heavy quarks to obtain the $p_T$ distribution of the 
$D$ and $B$ mesons. The following 
three sets of fragmentation functions have been used
to check the sensitivity of the results:

1. SET-I~\cite{seti}

\be
f(z) \propto \frac{1}{z^{1+r_Qbm_Q^2}}(1-z)^a exp(-\frac{bm_T^2}{z})
\ee
where $m_Q$ is the mass of the charm quark,
$r_Q$ =1, $a=5, b=1$ and $m^2_T = m^2_Q +p^2_T$. 
It has been explicitly checked that the $R_{AA}$ is not 
very sensitive to the values of $a$ and $b$.

2. SET-II~\cite{setii}
\be
f(z) \propto z^\alpha(1-z)
\ee
where $\alpha=-1$ for charm quark and it is $9$ for bottom\\

3. SET-III~\cite{setiii}
\be
f(z) \propto \frac{1}{\lbrack z \lbrack z- \frac{1}{z}- \frac{\epsilon_c}{1-z} \rbrack^2 \rbrack}
\ee
for charm quark $\epsilon_c=0.05$
and for bottom quark $\epsilon_b=(m_c/m_b)^2\epsilon_c$.

\begin{figure}
\begin{center}
\includegraphics[scale=0.43]{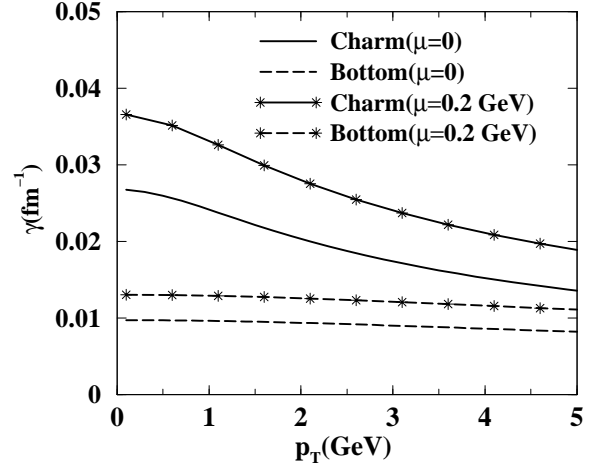}
\caption{Variation of drag coefficient with $p_T$ for $T=200$ MeV}
\label{fig5}
\end{center}
\end{figure}

\begin{figure}
\begin{center}
\includegraphics[scale=0.43]{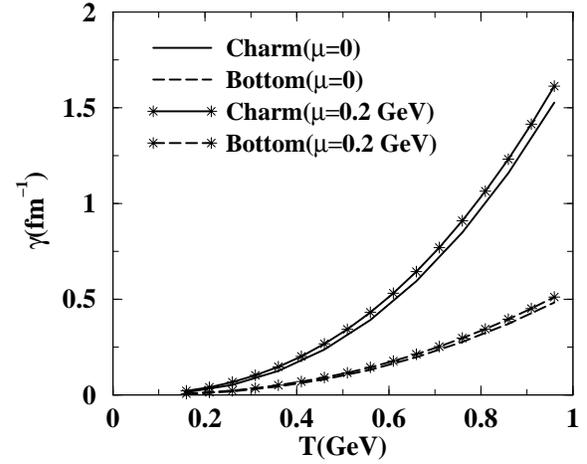}
\caption{Variation of drag coefficient with temperature.}
\label{fig6}
\end{center}
\end{figure}

\begin{figure}
\begin{center}
\includegraphics[scale=0.43]{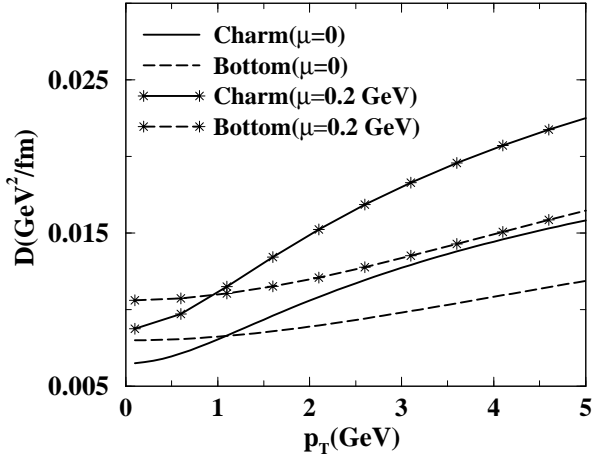}
\caption{Variation of diffusion coefficient with $p_T$ for $T=200$ MeV.}
\label{fig7}
\end{center}
\end{figure}

\begin{figure}
\begin{center}
\includegraphics[scale=0.43]{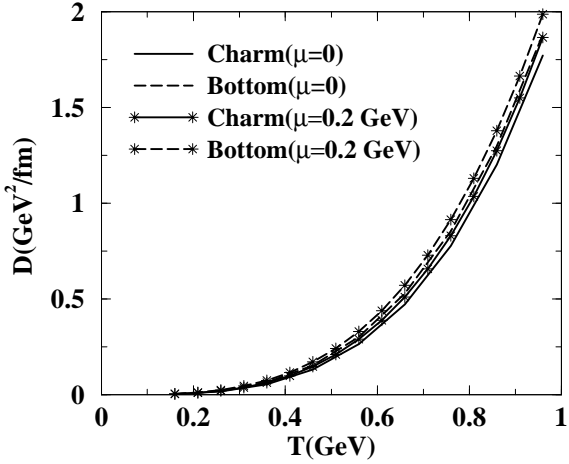}
\caption{Variation of diffusion coefficient with Temperature.}
\label{fig8}
\end{center}
\end{figure}


Recently, the $p_T$ spectra of $D$ mesons has been measured by the
STAR collaboration~\cite{starD} in Au+Au collisions at $\sqrt{s_{NN}}=200$ 
GeV. The $p_T$ spectrum of hadrons can be written as~\cite{ruuskanen}
\bea
\frac{dN}{d^2p_Tdy}&=&\frac{g}{(2\pi)^3}\int \tau r d\phi d\eta
(m_Tcosh(\eta-y)dr\nn\\
&&-p_Tcos\phi d\tau)f(u^\mu p_\mu)
\label{pvr}
\eea
$\eta$ is the space time rapidity, $p^\mu$ is the four momentum
and $u^\mu=\gamma(1,\beta)$ is the hydrodynamic four velocity,
$u^\mu p_\mu$ is the energy of the hadrons in the co-moving
frame of the plasma and  
$f(u^\mu p_\mu)$ is the momentum space distribution. 
In the spirit of the blast wave method we can write Eq.~\ref{pvr}
as~\cite{blastW}:
\bea
\frac{dN}{d^2p_Tdy}&=&\frac{g}{(2\pi)^3}\int \tau r d\phi d\eta\nn\\
&&m_Tcosh(\eta-y) f(u^\mu p_\mu) dr
\label{bw}
\eea
Taking the surface velocity profile as:
\be
\beta(r)=\beta_s\left(\frac{r}{R}\right)^n
\ee
and choosing n=1, the $p_T$ spectra of $D$ mesons is
evaluated.

Before comparing the data with non-equilibrium momentum distribution
we analyze the data within 
the ambit of the blast wave method~\cite{blastW} assuming a equilibrium
distribution for the $D$-meson. 
The values of the blast wave parameters {\it i.e.}
the radial flow velocity at the surface, $\beta_s$ and the freeze-out temperature, $T_{F}$ are
0.6   and 0.170 GeV respectively. The data is reproduced well~(Fig.\ref{fig9}). 
The value of $T_F$
is close to $T_c$, which indicate that the $D$-mesons (even if the
charm is in equilibrium in the partonic phase) can not
maintain it in the hadronic phase. This is reasonable because of the low
interaction cross sections of the $D$ mesons with other hadrons. 
Next we replace the
equilibrium distribution in Eq.~\ref{bw} by the solution of
FP equation appropriately boosted by the radial velocity. 
The results are displayed in Fig.~\ref{fig9}. The data is reproduced
well for all the three sets of fragmentation functions mentioned before.
The value of the freeze-out temperature is 170 MeV and the
flow velocity at the surface is 0.45, 0.35 and 0.4 for SET-I, SET-II and 
SET-III
fragmentation functions respectively. The values of $\beta_s$ is lower here
than the equilibrium case for all the fragmentation function. It is interesting to note 
that at low $p_T(\leq 0.5$ GeV) domain the results for equilibrium distribution 
substantially differ from the non-equilibrium distribution for
all the three sets of fragmentation functions. Therefore, measurements
of the heavy meson spectra 
at low $p_T$ domain will be very useful to distinguish between the equilibrium and
the non-equilibrium scenarios. The two scenarios also give different 
kind of variation at large $p_T$. The $p_T$ integrated quantity, {\it i.e.}
the $D$ meson multiplicity may also be useful to understand the difference
between the equilibrium and non-equilibrium scenarios.

\begin{figure}
\begin{center}
\includegraphics[scale=0.43]{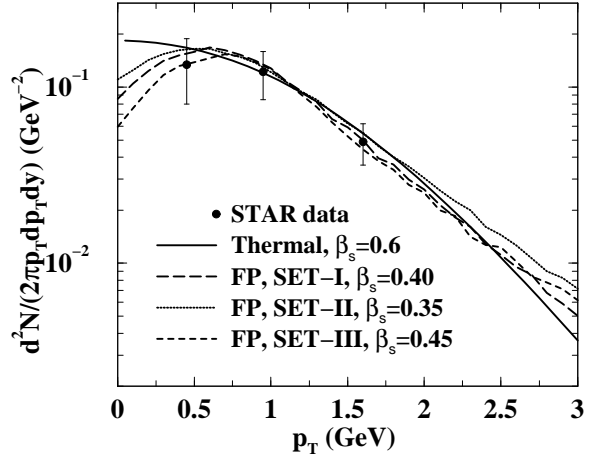}
\caption{Invariant momentum distribution of the $D$-meson as a function of $p_T$.}
\label{fig9}
\end{center}
\end{figure}

\section{Non-photonic single electron from heavy flavours.}

\begin{figure}
\begin{center}
\includegraphics[scale=0.43]{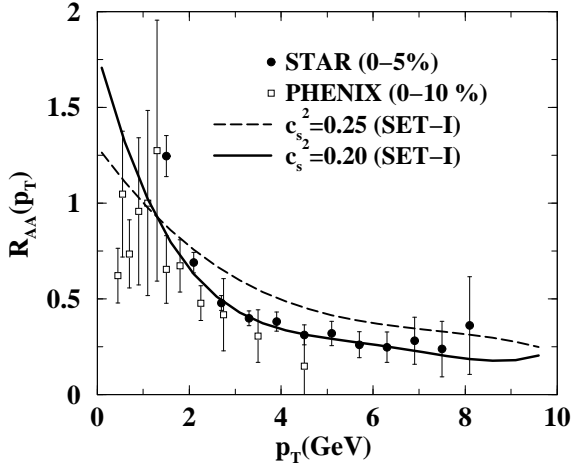}
\caption{Nuclear suppression factor, $R_{\mathrm AA}$ as function of $p_T$.}
\label{fig10}
\end{center}
\end{figure}

\begin{figure}
\begin{center}
\includegraphics[scale=0.43]{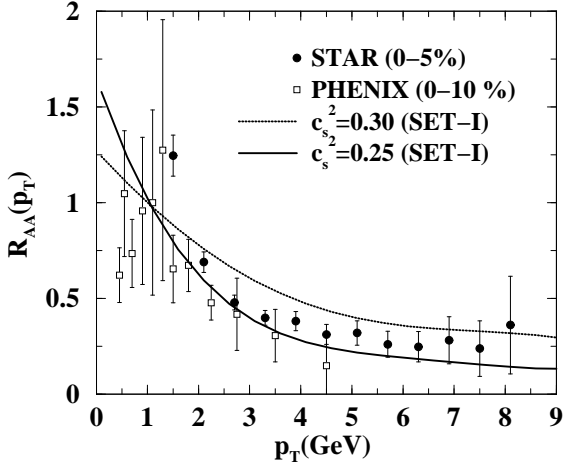}
\caption{Same as Fig.~\protect{\ref{fig10}}
with enhancement of the cross section by a factor of 2.}
\label{fig11}
\end{center}
\end{figure}

\begin{figure}
\begin{center}
\includegraphics[scale=0.43]{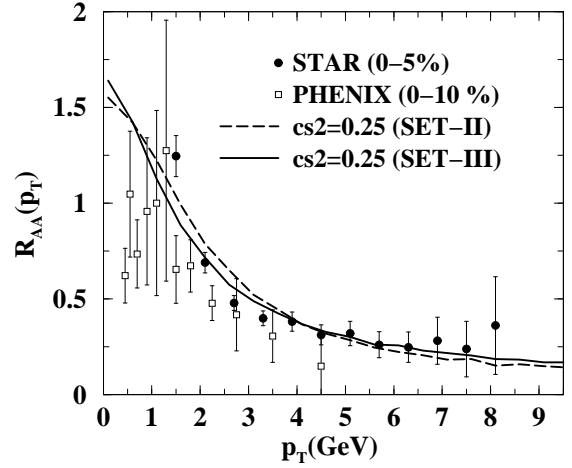}
\caption{Same as Fig.~\protect{\ref{fig11}}
for fragmentation functions of SET-II and III.}
\label{fig12}
\end{center}
\end{figure}

The STAR~\cite{stare} and the PHENIX~\cite{phenixe} collaborations have
measured the non-photonic single electron inclusive $p_T$ spectra recently
both for Au+Au and p+p collisions at $\sqrt{s_{NN}}=200$ GeV. The quantity 
\be
R_{AA}(p_T)=\frac{\frac{dN^e}{d^2p_Tdy}^{\mathrm Au+Au}}
{N_{\mathrm coll}\times\frac{dN^e}{d^2p_Tdy}^{\mathrm p+p}}
\label{raa}
\ee
called the nuclear suppression factor, 
will be unity in the absence of any medium. However, the experimental
data from both the collaborations~\cite{stare,phenixe}
shows substantial suppression ($R_{AA}<1$) for $p_T\geq 2$ GeV indicating
the interaction of the plasma particles with the charm and bottom quarks 
from which electrons are originated through the process:
$c(b)$ (hadronization)${\longrightarrow}$ $D(B)$(decay)$\longrightarrow$
$e+X$.  The loss of energy of high $p_T$ heavy quarks propagating through
the medium created in Au+Au collisions causes a depletion of high $p_T$
electrons. 

To evaluate $R_{\mathrm AA}$ theoretically, the solution of the FP equation 
for the charm and bottom quarks should be convoluted by the fragmentation functions
to obtain the $p_T$ distribution of the $D$ and $B$ mesons which subsequently
decay through the processes:  $D\rightarrow X e \nu$ and 
$B\rightarrow X e \nu$. Similar formalism has been used in~\cite{rma}
to study the evolution of light quark momentum distributions.
The resulting electron spectra from the decays of $D$ and $B$ mesons 
can be obtained as follows~\cite{gronau,ali,akc}:
\be
\frac{dN^e}{p_Tdp_T}=\int dq_T \frac{dN^D}{q_Tdq_T} F(p_T,q_T)
\ee
where 
\be
F(p_T,q_T)=\omega\int \frac{d(\bf{p}_T.\bf{q}_T)}{2p_T\bf{p}_T.\bf{q}_T}g(\bf{p}_T.\bf{q}_T/M)
\ee
where $M$ is the mass of the heavy mesons ($D$ or $B$), 
$\omega=96(1-8m^2+8m^6-m^8-12m^4lnm^2)^{-1}M^{-6}$ ($m=m_X/M$) and $g(E_e)$ is given by
\be
g(E_e)=\frac{E_e^2(M^2-M_X^2-2ME_e)^2}{(M-2E_e)}
\ee
related to the  rest frame spectrum for the decay $D(B)\rightarrow X e \nu$
by the following relation~\cite{gronau}
\be
\frac{1}{\Gamma}\frac{d\Gamma}{dE_e}=\omega g(E_e).
\ee

We evaluate the electron spectra from the decays of heavy mesons
originating from the fragmentation of the heavy quarks propagating 
through the QGP medium formed in heavy ion collisions.
Similarly the electron spectrum from the p-p collisions
can be obtained from the charm and bottom quark distribution
which goes as initial conditions to the solution of FP equation.
The ratio of these two quantities gives the nuclear suppression 
$R_{AA}$. For a static system the temperature dependence of the drag and
diffusion co-efficients of the heavy quarks enter via the
thermal distributions of light quarks and gluons through
which it is propagating. However, in the present scenario
the variation of temperature with time is governed by
the equation of state or velocity of sound 
of the thermalized system undergoing hydrodynamic 
expansion. In such a situation the quantities like $\Gamma$ (Eq.~\ref{cgama})
and hence $R_{\mathrm AA}$ becomes sensitive to velocity of sound 
in the medium.

The results for $R_{AA}$ is displayed in Fig.~\ref{fig10}. 
The theoretical results are obtained for fragmentation function
of SET-I~\cite{seti}. The velocity of sound for the QGP
phase is taken as $c_s=1/\sqrt{4}$  corresponds to the 
equation of state, $p=\epsilon/4$. The results failed to describe
the data in this case. Next we generate $R_{AA}$ by changing the
value of $c_s$ to $1/\sqrt{5}$ and keeping all the other parameters
fixed. The resulting spectra describes the data reasonably well. 
Lower value of $c_s$ makes the expansion of the plasma slower
enabling the propagating heavy quarks to spend more time to interact
in the medium and hence lose more energy before exiting from the plasma
which results in less particle production at high $p_T$. Further lowering
of $c_s$ gives further suppressions.

However, as mentioned earlier the non-perturbative effects are
important for the interaction of the heavy quarks with
the plasma. Therefore, we enhance the cross section by a factor
of 2 and find that  
the experimental results can also be explained by taking an EoS $P=\epsilon/4$
(~Fig\ref{fig11}) and keeping other quantities like
fragmentation functions etc unchanged. 
The ideal gas EoS $P=\epsilon/3$ can not reproduce the data
even if the cross section is enhanced by a factor of 2.
With  $c_s=1/4$ and enhanced cross section (by a factor 2)
the data can also be described with for
fragmentation functions of set-II and set-III~Fig.~\ref{fig12}.
Several mechanisms like inclusions of non-perturbative
contributions from the quasi-hadronic bound state~\cite{hvh},
3-body scattering effects~\cite{ko} and employment of running coupling
constants and realistic Debye mass~\cite{gossiaux}
have been proposed to improve the description of the  experimental data.
It is demonstrated here that the EoS of the medium and the
non-perturbative effects  play a crucial
role in determining the nuclear suppression factor.

\section{Summary and Conclusions}
The transverse momentum spectra of the $D$ and $B$  mesons 
have been studied within the ambit of Fokker-Planck equation
where the charm and bottom quarks are executing Brownian 
motion in the heat bath of light quarks and gluons.
We have evaluated the drag and diffusion co-efficients
both for zero and non-zero quark chemical potential.
Results for non-zero baryonic chemical potential
will be very useful for studying physics at low energy RHIC run~\cite{santosh}.
The results are compared with experimental
data measured by STAR collaboration. It is found that the present experimental
data can not distinguish between the $p_T$ spectra obtained from the
equilibrium and non-equilibrium charm distributions. Data at lower
$p_T$ may play a crucial role in making the distinction 
between the two. Since the results for equilibrium and
non-equilibrium scenarios differ, the $p_T$ integrated quantity {\it i.e.}
the $D$ meson multiplicity may also be very useful to put
constraints on the model. The nuclear suppression factor for the measured
non-photonic single electron spectra resulting from the semileptonic decays of
hadrons containing heavy flavours have been evaluated using the present
formalism.  The experimental data on nuclear suppression
factor of the non-photonic electrons can be reproduced within this
formalism if the expansion of the bulk matter is governed by a equation 
of state $p=\epsilon/4$ and the partonic cross sections are
taken as $2\times\sigma_{\mathrm pQCD}$.
Three different kinds of fragmentation functions for the
charm and bottom quarks hadronizing to $D$ and $B$ mesons
respectively have been used and found that $c_s\sim 1/\sqrt{4}$
can describe the data reasonably well. The data can not be reproduced
with $c_s\sim 1/\sqrt{3}$ even after enhancing the cross section by a factor
of 2. The loss of energy by the heavy quarks due to radiative process
may be suppressed due to dead cone effects. In the present work
the radiative loss is neglected.
The FP equation
needs to be modified to include the  radiative loss~\cite{pqm,glv,zoww,bdps,
salgado}
(see ~\cite{revieweloss} for a review),
work in this direction is in progress~\cite{santosh}.

The calculations may be improved by making 
the space time evolution picture more rigorous as follow.
In the absence of any external force the evolution of the heavy quark 
phase space distribution is governed by the equation: 
\bea
\left (\frac{\partial}{\partial t}+{\bf v_p\cdot\nabla_r}\right)
f({\bf p,r},t)=C[f({\bf p,r},t)]
\label{eq3d}
\eea
As mentioned before the FP equation can be obtained from
the above equation by linearizing the collision term, C[f({\bf p,r},t)].
To take into account the energy loss of the heavy quarks in
the thermal bath the ideal hydrodynamic equation needs to be modified as:
\be
\partial_\mu T^{\mu\nu}=J^\nu
\label{hydrosource}
\ee
containing a source term $J^\nu$ corresponding to the energy-momentum 
deposited in the thermal system along the trajectory of the heavy quark, 
which may be taken as $J^\nu\sim dp^\nu/d\tau$~\cite{akcheinz} ($p^\nu$ is
the four momentum vector).
Eq.~(\ref{hydrosource}) should solved for  $T({\bf r},t)$ 
with appropriate equation of state which
can be used to obtain the surface of hadronization by setting 
$T(\bf{r_c},t_c)=T_c$. Subsequently the solution of
Eq.~\ref{eq3d}, $f({\bf r_c},t_c)$ for the heavy quark 
on this surface should convoluted with the fragmentation
function to obtain the $B$ and $D$ distribution.

{\bf Acknowledgment:} We are grateful to Bedanga Mohanty and Matteo Cacciari
for very useful discussions. We also thank Jajati K Nayak for his involvement
in this work initially. This work is supported by DAE-BRNS project Sanction No.
2005/21/5-BRNS/2455.

\end{document}